# Power Networks: The Digital Approach

Camila Fukuda[1], Henrique Pita[1], Roberto Rojas-Cessa[2] and Haim Grebel[1*]

ECE Department, the New Jersey Institute of Technology, Newark, NJ 07102

[1] also, The Electronic Imaging Center at NJIT, Newark, NJ 07102

[2] also, The Networking Research Laboratory, NJIT, Newark, NJ 07102

*grebel@njit.edu

**Abstract** – In the current analog grid, power is available at all times, to all users, indiscriminately. This makes the grid vulnerable to demand fluctuations and much effort has been invested to mitigate their effect. The Digital Power Network (DPN) is an *energy-on-demand* approach to power grids. In the new (digital) approach, the user initiates an energy request. This action alleviates uncertainties in energy demands. The service provider may grant the request fully or partially. Energy is then transmitted in discrete units, analogous to packets of data over a computer network. The packetized energy is routed to the user's address. Because energy demands are known ahead of time, the energy provider may optimize the power distribution of the entire power network and isolate pockets of instabilities. For example, under severe energy constraints the energy provider may queue some energy requests and grant these requests later. Alternative energy resources may be seamlessly incorporated into the power network as yet another address in the system and since their energy is coded, they would be connected to specific users directly. In its simplest form, this grid can be realized by overlaying an auxiliary (communication) network on top of an energy delivery network (the current transmission lines) and coupling the two through an array of addressable digital power switches. Optimization of energy requests is the topic of this paper. We investigate the role of the network queue and provide a snapshot of its behavior in time. DPN with a limited channel capacity and the optimal path for energy flow in a standard IEEE 39 bus are considered, as well.

## I. A Short Introduction

The underlying concepts of the current power grid have remained unchanged during the past 150 years [1-2]. Increased grid monitoring and heavy investment in predictive models are often the response to increasing grid's complexity [3-6]. These upgrades, while important, cannot sustain the grid through new challenges, such as the seamless integration of sustainable energy sources [7]. Integration of alternative energy resources with the power grid is of global interest due to the desire to reduce reliance on fossil fuels and the incorporation of environmental friendly energy resources [8]. Currently, it takes a tremendous effort to integrate alternative energy sources into the power grid since the grid's stability could be largely compromised by these sources. An increasingly larger number of systems directly and indirectly contribute to the welfare of the energy distribution systems and the breakup of one affects many others.

The original design of the power grid was to transport power in a single direction, from the generator to the loads (the users, or the customers). It is a simple distributive network available to all – any user may connect to it at any time at his/her own discretion. In contrast, information networks (or computer networks) allow bi-directional communication and negotiate the transfer of information upon request. Computer networks rely on some or all of the following: memory devices, system addresses, network of routers (or smart power switches) and established protocols. In computer network terms, the power grid is but the physical layer of the network and our vision is to add management and control plane to it. In the digital approach to the grids, energy distribution is managed up to the level of the user's address. The immediate question would be – who needs it? The simple answer is that retroactive response to developing emergencies makes the current grid vulnerable to power fluctuations, which ultimately lead to blackouts.

Here, we describe a wholly power network concept – a digital power network in which power is transmitted and delivered in packets similarly to computer networks [9-11]. We remind ourselves that computer networks carry energy and obey Kirchhoff's laws similarly to the current power grid; it is the modulation of energy (the bit) that conveys information. Power packets have boundaries and destination addresses. Power packets are delivered to particular users through a path determined by smart power switches and in specific amounts according to a request-grant protocol.

So the next question would be: why don't we use information systems to deliver energy? The answer is that information systems cannot simply be mapped onto the power grid setting. While the name digital grid (or, smart grids) have been mentioned in the past, in reality, many of the proposed schemes were not scalable, would not take into account the random nature of energy requests (amount and time), and suffered from poor adaptation of communication systems (characterized by low-voltages, low-currents at ultra-high frequencies) to power networks (characterized by high-voltages, high-currents at super-low frequencies). It is the large random variations in requested current levels that makes the mapping of presently deployed information networks onto the power world so difficult. We comment on DPN realization at the end of this paper.

As is the case for the transmission of data packets throughout the Internet, various protocols are adopted and their functions are well defined. Similarly, users of power networks initiate the delivery of energy packets by issuing requests in the *transport layer* for the amounts and duration of needed power. The utility provider then grants these requests, fully or partially based on the availability of power in the grid. Price and amounts can then be negotiated between providers and users. In the *control plane*, routers and power switches enable the energy flow toward the users along efficient paths.

## II. The Digital Power Network (DPN)

The DPN uses a demand-supply management model [9-12]. Users (or loads) issue a request for the amount of energy in demand, and the service provider allocates energy to the selected users at any given time. An immediate advantage to this approach is that energy allocation may be optimized in real time (namely, delayed or advanced) depending on the overall grid status. Owing to the request-grant allocation protocol, energy demands and energy consumptions may be closely monitored and safety margins can be optimized for any given moment, thus increasing the overall power network's efficiency. During the recovery of power networks from a blackout, the energy provider can isolate pockets of vulnerabilities by avoiding delivery to their addresses. The approach is wholly and considers every aspect of the network: power generation, distribution, and usage.

Fusing data and energy [13-16] *in discrete formats* dramatically reduces management complexity because, in principle, the energy (power delivered over time) can be directed to specific users. The digitation of time (through allocation of energy at varying time slots), or the digitation of power (to be delivered by discrete current levels while keeping the voltage constant) are two possible digitization approaches. The simplest adaptation of the DPN is by interfacing the power network with an auxiliary data network that opens and closes power switches along the way as the energy is routed to the addressed users (Figure 1(a)). The energy supplier selects the appropriate smart load through data fused to the energy packet. Figure 1(b) shows the synergistic operation of data and energy where a data network provides the management and control of the power network. An alternative energy source (e.g., a solar panel) may be incorporated into the distribution loop as yet another address (see path allocation below). The data network manages the communications between users and the distribution point, energy delivery, and the management of the two energy sources at a much higher frequency than the frequency used to transmit the electrical power itself.

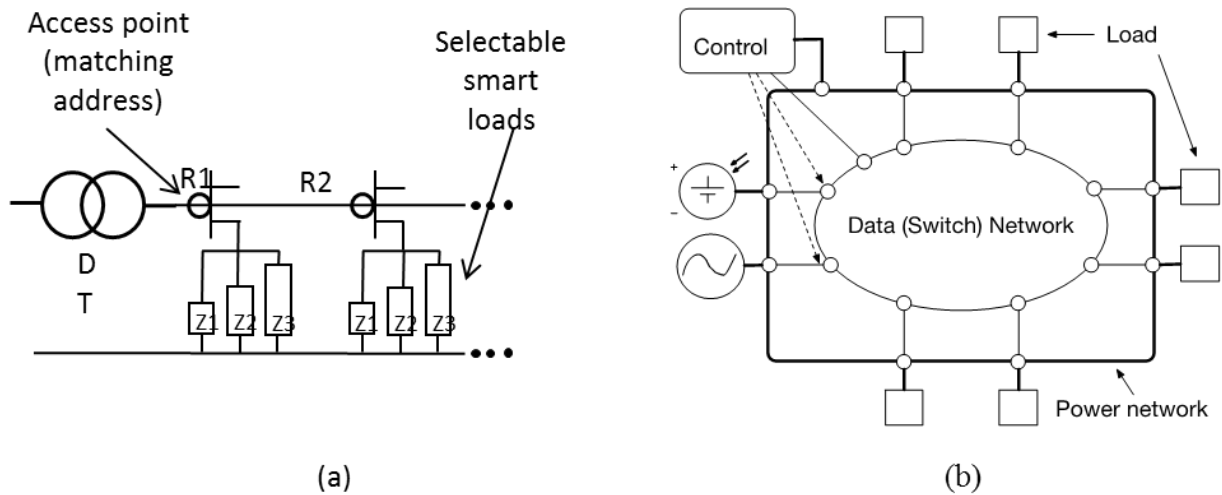

*Figure 1. (a) Smart limiters and (b) the simplest framework of two overlaid networks: the data (switch) network, which is coupled to the power network through controlled smart loads. Also shown are AC and solar power sources.*

While the concept of digital power networks seems straight forward, its implementation is not. For example, what is the role of memory (energy storage)? How would you configure bi-directional energy transfer? What is an energy bit looks like? How does an energy packet look like? What is the penalty for the information overhead? The focus of this paper is on the optimization of energy delivery. Current power-grid models and simulations tools are not fully equipped with the concept of a control plane and need to be modified. We instead, run statistical models to answer the following question: how many request cycles are needed before a customer is satisfied under limited power conditions? Answering this question will help us design the minimum energy cap for DPN.

## II.1. Latency and Storage Issues

For simulation purposes users are requesting energy, randomly. We assign a probability to users who request energy to turn their appliance ON (meaning they start with an appliance OFF) and another probability for those users who have their equipment already ON and wish to continue to do so. An example is provided in Figure 2: two random numbers are generated for each user. For those users that were OFF in the previous round, we check whether the randomly generated number, $p_{req}$, is smaller than a given request probability, $p_{request}$. If yes, then a new random number is generated for the actual energy request. The second randomly generated number is used for those users whose equipment is already ON. If the randomly generated number, $p_{on}$, is larger than $p_{stay\_on}$, then their new request will be 0 (they will be turned OFF). Otherwise, they will remain ON with their previous energy request. Unsatisfied energy requests are sent to the queue. One could generate a single random probability number and compare it to the $p_{request}$ and $p_{on}$ for the two groups involved as in step 2 of Figure 2: the group with its power ON in the previous step and the group with its power OFF in the previous step. Both approaches yielded similar results. Finally, we note that the process is not entirely Markovian; the queue memorizes the size and the order of the requested energy until satisfied or until the request is dropped. Specifically, a 2-state Markovian chain is an adequate analytical model for ON and OFF states as long as the queue is empty. A three-state Markovian chain has a 6% discrepancy with the numerical results because the queue has a selection rule and does not accept the energy requests randomly. Specifically, we considered two examples: satisfy the large energy requests first (hence sending the reaming small amounts to the queue) and separately, satisfy small energy requests first (hence sending the large energy requests to the queue).

Step 1: (ON=logical(e_requested))

preq=rand(1,numel(e_requested));              %Generate number to compare with p_request
pon=rand(1,numel(e_requested));               %Generate a number to compare with p_stay_on

Step 2:
e_requested(not(ON) & preq<p_request)=rand(1,numel(e_requested(e_requested==0 &
preq<p_request)));                            %Generate a random request when turning on
e_requested(ON & pon>p_stay_on)=0;            %Turn off

*Figure 2. An example of simulating steps in MatLab for the Digital Power Networks.*

Current grids do not use an energy cap although they are limited by the overall generated power. Local array of fuses provide additional protection. Predictive models and years of data collection aid the utilities in forecasting the level of service. On the other hand, micro-grids, such as a house equipped with a small power generator, are limited by the generator's capacity. So in our simulations we set an energy cap for the total energy available per each cycle of energy request (round). As the two probabilities become larger, more users (or for that matter, addresses of appliances) turn their equipment ON and more users whose equipment is already ON remain ON. That situation puts an unsustainable burden on the power network - the power network cannot satisfy all users at the same time and unsatisfied requests are sent to the queue. In the current analog grid, the system will become overloaded and fail.

How long or how many grant-cycles will a certain user wait in the queue until the request is satisfied? The answer depends on the probability that an energy request of a particular user will stay in the queue. In extreme cases, the users would stay in the queue until satisfied. A finite probability to stay in the queue provides us with the freedom to let the user choose a limited time frame during which energy is still needed. Thus, if that probability to stay in the queue is small, then the energy request will eventually be dropped from the queue and the number of remaining requests could be satisfied more quickly. If, on the other hand, the probability of staying in the queue is large (probability 1 for staying in the queue until the request is satisfied), then the waiting period may be prolonged. If the energy cap is large enough and all requests are satisfied, then there will be no queued requests. In this case, the DPN behaves similarly the current power grid, yet with a direct knowledge of the grid's status during the request cycle and full control over the energy flow (Figure 3,4).

How do we decide who is to receive energy and whose request is to be sent to the queue? There are two approaches that can be employed. In the first, and the one used to generate Figures 3-4, we satisfy the smaller energy requests first and send the overflow of requests to the queue. The second approach, used to generate Table 1 and Figure 5, was to satisfy the largest energy requests first and send the overflow requests to the queue. The two approaches would result in different queue time and number of requests waiting to be satisfied. Satisfying the smaller energy requests first, would results in smaller number of queued and large energy requests albeit with prolonged waiting periods.

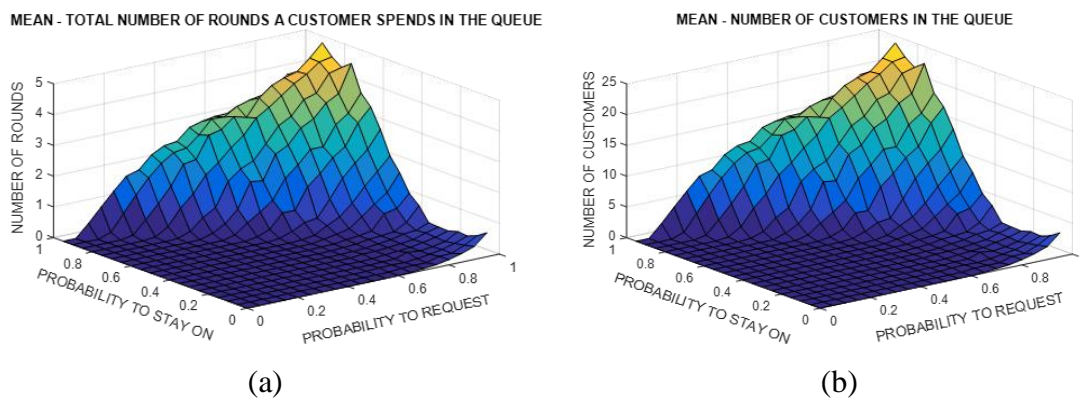

(a) (b)

*Figure 3. (a) Mean numbers of rounds and (b) users (customers) waiting in the queue when the probability of waiting in the queue is 0.1. Here, we consider a total number of users to be 500. The channel capacity (energy cap) was set to 150 units of energy and each user could ask for up to one unit of energy.*

When the probability of staying in queue increases to 0.5, the waiting period (in number of rounds) and the number of users waiting in the queue would obviously increase, as shown in Figure 4. While the probability to stay in the queue increased 5 times, the maximum number of cycles in the queue has not increased as much; the DPN was able to distribute the requests quite efficiently.

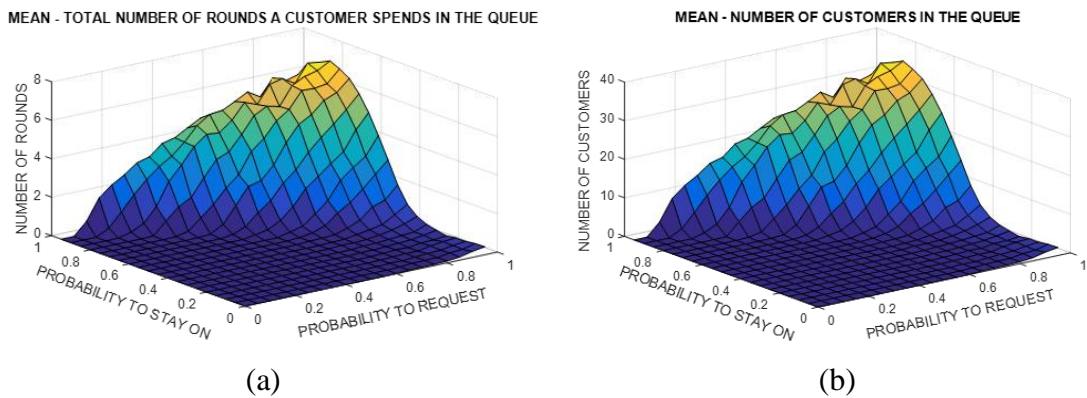

*Figure 4. (a) Mean of the total number of rounds that a user spends in the queue and (b) the mean number users waiting in the queue, when the probability of waiting in the queue is 0.5. This means that more users are waiting in the queue compared to the case presented in Figure 4. The total number of users is 500 and the cap is set to 150 units, where each user may ask for up to one unit of energy.*

Having the ability to schedule the delivered energy makes us wonder if adding a battery to the DPN would decrease the waiting time for a waiting/demanding user. The short answer is no - even with a very large battery, the slow charge/discharge cycles may not cope with the fast and randomly distributed demand fluctuations. As we have shown earlier [11] one requires a minimum time of a least 1 minute to meaningfully charge a laptop battery. However, if every customer is equipped with a supercapacitor/battery system, and if all of these energy storage units are at the service of the entire grid (namely, they may deliver energy to other users, as well), then one may imagine that not only we may achieve fast charge/discharge cycles but also a more regulated grid. Such approach, which we dub '*cloud energy storage*,' has the effect of increasing the energy cap for the power network. While fast charging and discharging of large amounts of electrical energy make supercapacitors ideal for short-term energy storage the amount of energy stored is rather limited with today's technology. The DPN takes a wholly approach to storage and delivery of energy on an energy cycle basis.

What will happen if the energy cap is increased to 2/3 of the maximum users' requests? Surprisingly, the digital grid can accommodate all random requests without the need for a battery and without sending users' requests to the queue. Our simulations showed that batteries will be charged and will stay charged all the time, independently of the probabilities of staying ON or OFF.

Example: consider an office building where all occupants turn on their air-condition units at the same time. As the units are turned ON, they draw up to 8 times more power than their steady state levels. The DPN protocol would then delay some of these random requests for no significant impact on service but with a significant energy saving. The example may be scaled down to a single residential home.

Our simulations take into account random energy requests for each customer (Figure 5). This is a simple case of a power grid with capacity of 250 energy units and 500 customers. Each round presents one cycle of energy requests (say every 0.5 second). For most cases, the DPN can satisfy all users except for the case where the probability to stay ON approaches 1. Thus, for most scenarios (probabilities) the situation is very similar to the present grid. *However, and unlike the present grid, during extreme cases, the overflow of users' demands is placed in a queue and would not overwhelm the entire grid, thus avoiding black outs.*

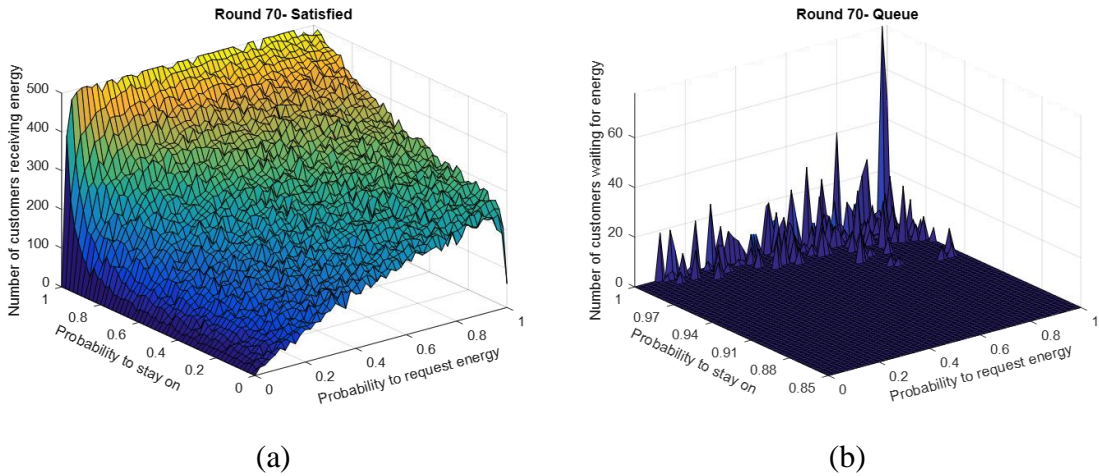

(a)    (b)

*Figure 5. (a) In most cases the digital grid accommodates all energy requests. When the demand is large (probability to stay ON is close to 1) some users are sent to the queue. (b) Number of users waiting in the queue. The scale is focused on the large probability range for staying ON.*

Another example is a snapshot of one time slot (one request cycle, or one round) of a 10-user digital micro-grid (Table 1). These 10 users may tap into an energy storage while their energy grant is pending. Each user may consume *no more* than one unit of energy (but could consume less). Let us consider an energy cap on the energy consumption of 3 energy units. The probability to switch from OFF to ON and to continue to request a service are both 0.3. These switching probabilities are fairly low but the energy cap is also very low; it is 30% of the maximum consumed energy by all users. The maximum stored energy in the battery is set to 10% of the maximum total energy (1 unit in total) that may be consumed in one cycle. Let us consider Users 2 and 7 in Table 1. The users request various amounts of energy at some point in time. Because the energy cap is only 3 units of energy, the system cannot accommodate all requests and some users are put on hold (namely, their requests are queued). Priority in this case was given the largest energy requests in reverse order thus the smallest request was queued. In order to avoid long delays in the energy supply, User 2 taps into a battery resource. The energy storage may be physically situated on the user premise or shared amongst all users (*cloud energy storage*). If there is an excess of energy in the system, then the energy storage is charged. Fast charge/discharge of the energy storage is required because the requests are varying for each round.

| Users: | 1 | 2 | 3 | 4 | 5 | 6 | 7 | 8 | 9 | 10 | Total |
|---|---|---|---|---|---|---|---|---|---|---|---|
| Request | 0.4974 | 0.4869 | 0 | 0.5473 | 0 | 0 | 0.5221 | 0 | 0.9519 | 0 | 3.0056 |
| Grant | 0.4974 | 0 | 0 | 0.5473 | 0 | 0 | 0.5221 | 0 | 0.9519 | 0 | 2.5187 |
| Queued | - | 1 | - | - | - | - | - | - | - | - | 1 |
| Storage | 0 | 0.4869 | 0 | 0 | 0 | 0 | 0 | 0 | 0 | 0 | 0.4869 |

*Table 1. A snapshot of requested, granted, queued, and stored energy for a 10-user micro-grid network.*

## II.2. Optimization Scenarios – Energy Storage

Here, the energy storage acts as a secondary energy source, and the power network provides for the primary source of energy. The energy storage is treated as an addressed user when extra energy, left by the optimization process, and energy is routed in its direction. The energy storage stores up to 10% of grid's capacity. Let us also consider that only designated users, which typically is set as 10% of the total number of users, are allowed to tap the energy storage. When the primary source cannot offer energy to the users, they enter the queue and try to tap into the secondary source. These users do not leave the queue because there is no guarantee that the storage has energy during the next round. Below, we have used an optimization approach that minimizes the wait in the queue.

There are several ways to handle the energy requests, and thus, the optimization of the energy flow. At the present time, the (analog) power grid 'regulates itself', meaning that the loads determine the level of current consumed by the grid. In cases where the generator cannot deliver, blackouts occur. One family of optimization algorithms is the genetic algorithm. It is based on a learnt process (namely, collecting data for several request cycles), yet allows for random events.

### II.2.a. Genetic Algorithms

Genetic Algorithms optimize the energy allocation by using a few simple rules:

1. Selection: select parents from population (energy users) for the next generation of solutions (children). In our case we fit the incoming small energy requests first, moving on to the larger requests until we reach the channel capacity. The larger energy requests are sent to the queue. Obviously, instead one can accommodate the largest energy requests first, or any combination of the above. In a large pool of users and limiting the maximum demand to one energy unit the selection process does not significantly change the outcome.

2. Crossover: just like in biology, the characteristics of the parents in each generation are "mixed" into several possible solutions (population/children).

3. Mutation: random changes happen in the characteristics of the parents to generate the children.

The code is part of optimization package of Matlab. We considered the following experiment setup: number of users – 500; channel capacity – 200 units (maximum 1 unit per user); storage capacity of 20 energy units; the probability of a request to stay in the queue equals to 1; the probability to change status in the queue is equal to 0; and the number of preferred users that can tap into energy storage elements was 40. The results are shown in Figure 6.

The genetic algorithms are learnt algorithms; for random processes they reach optimization after several rounds and thus, the process was run for 100 or 50 cycles to reach an averaged solution and run for 20 more cycles to reach a higher degree of optimization. Time wise, it translated to a longer computing times. It took 1 sec to handle 500 customers to reach a non-optimized solution on a laptop computer. It took 283 sec for the genetic algorithm to reach a much better optimized energy distribution among 500 users while using the same laptop machine.

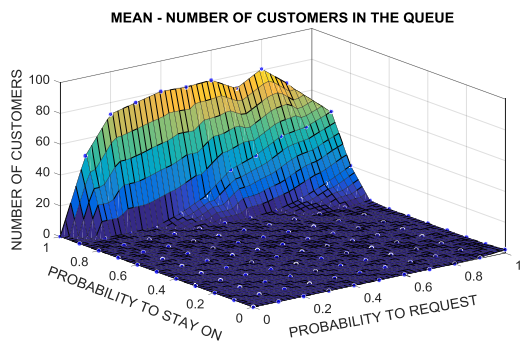
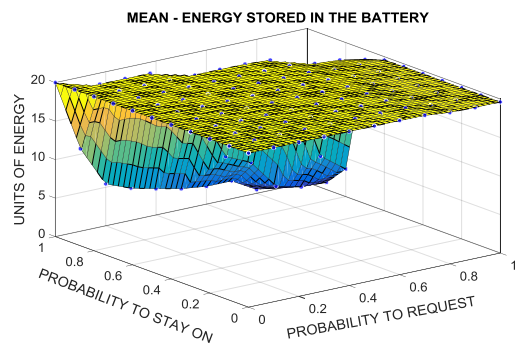

(a)           (b)

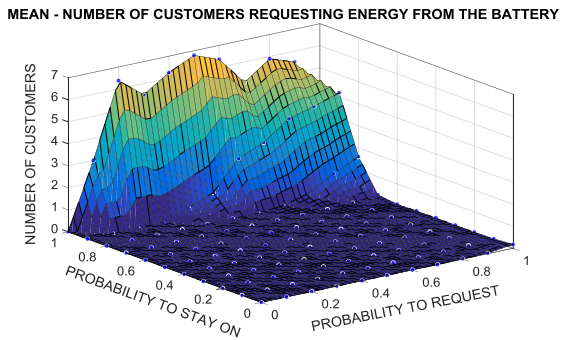
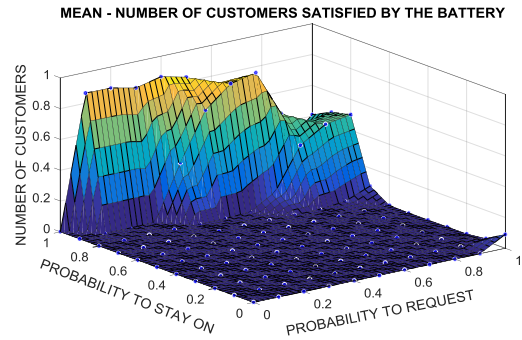

(c)           (d)

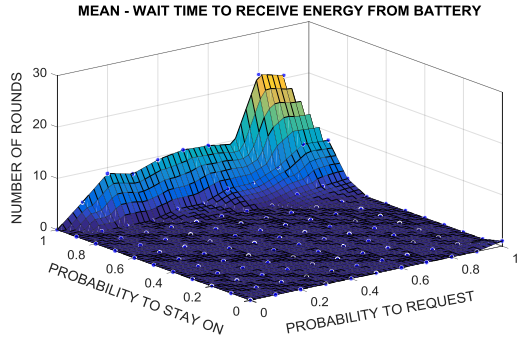
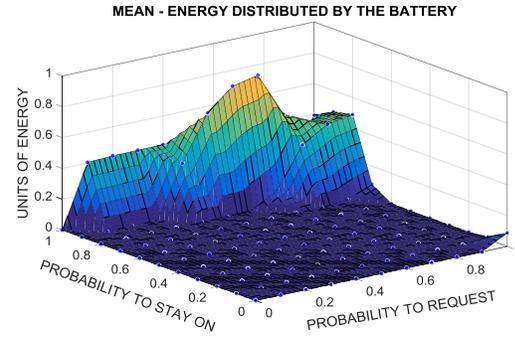

(e)  (f)

*Figure 6. Mean values for requested energy and satisfied users. In obtaining the mean values, the program was running for 20 simulations and collected information during 50 cycles (rounds) each. Only 40 users out of 500 users may tap into this additional resource.*

Since the requests are randomly changing from one energy request cycle to another, it is better to present both the mean and its standard deviation. In Figure 7 we show 500 users in a power network with a capacity of 200 energy units. The mean value well represents the situation at low probabilities. When the network reaches its full capacity, the variation in the number of users sent to the queue obviously increases.

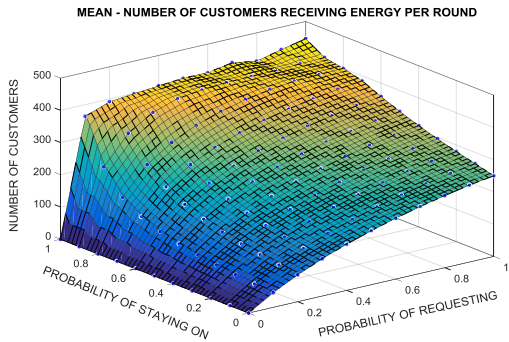
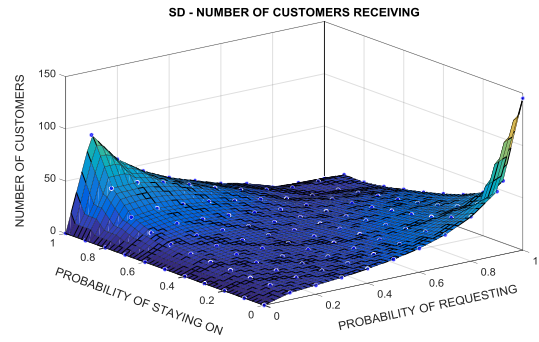

(a)  (b)

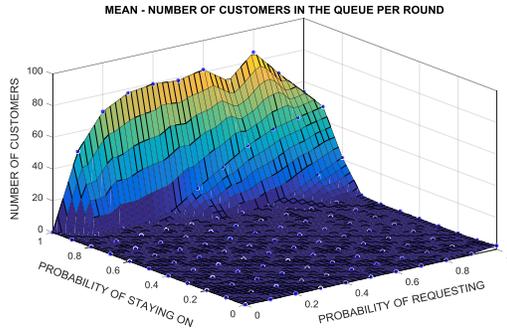 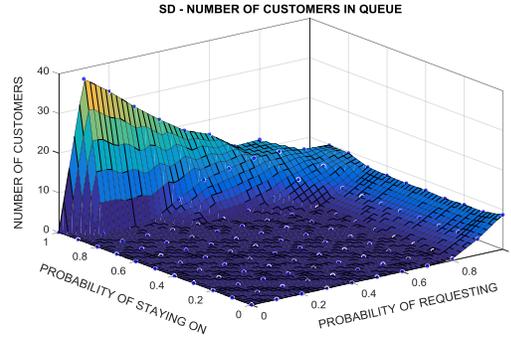

(c)                      (d)

*Figure 7. Mean values (a,c) and the standard deviation (b,d) for users receiving energy and in the queue. In obtaining the mean values, the program was running for 20 simulations and collected information during 50 cycles (rounds) each.*

A direct comparison for cases with and without the energy storage is made in Table 2 below. The simulations conditions are somewhat different: number of simulations – 50; number of cycles (rounds of time slots) – 50; number of users – 500; number of special users (only those can tap into the sustainable resource) – 50; energy cap (or channel capacity) – 100; the probability to stay in the queue – 1 (meaning no one leaves the queue unless satisfied); the probability to change your queue status – 0; battery capacity – 10 energy units. Each user may request up to 1 energy unit. The comparison was made for a probability to stay ON set to 0.5 and the probability to request energy if the user was at OFF state as 0.5.

Thus, the impact of the battery on the digital power network is relatively small; the energy storage element holds relatively small amount of energy and needs to charge part of the time. A larger impact is achieved if the battery is interfaced with a solar panel of equal energy capacity. The charging of the battery is not made on the expense of the grid and what extra solar energy left is added to the network energy capacity (Table 3).

|  | 50 SPECIAL CUSTOMERS | NO BATTERY |
|---|---|---|
| Energy distributed per round: | 99.5224 | 99.5326 |
| Energy requested per round: | 189.0862 | 190.4968 |
| Number of customers in the queue per round: | 91.5224 | 92.9716 |
| Number of customers that received energy per round: | 203.1692 | 202.5528 |
| Number of customers that requested energy per round: | 294.6916 | 295.5244 |
| Number of customers that entered the queue per round: | 6.7464 | 6.8888 |
| Number of customers that were satisfied in the queue per round: | 4.6908 | 4.8032 |
| Number of customers not satisfied by the end of the rounds: | 102.7800 | 104.2800 |
| Number of customers that were never satisfied in the queue: | 79.3600 | 79.2000 |
|  |  |  |
| Energy distributed by the battery per round: | 0.4662 | - |
| Energy available in the battery per round: | 9.4556 | - |
| Energy requested from battery per round: | 9.4556 | - |
| Number of customers that requested from the battery per round: | 9.2533 | - |
| Number of customers that received energy from the battery per round: | 0.4752 | - |
| Number of customer that were never satisfied by the battery: | 25.8200 | - |
| Total energy delivered (Battery+Grid) per round: | 99.9886 | 99.5326 |
|  |  |  |
| Number of rounds a customer in the queue waits to be satisfied: | 10.5674 | 10.7096 |
| Number of rounds a customer spends in the queue | 9.1522 | 9.2972 |
|  |  |  |
| Wait time to receive energy from the battery: | 9.4713 | - |
| Total Rounds spent requesting energy from battery: | 0.9456 | - |

*Table 2. A power network interfaced with a power storage (battery). The effect of the battery is quite small. Wait times are measured in cycles (rounds, or time slots). 'Energy requested per round' includes new and queued requests.*

## II.3. Optimization Scenarios – Solar Energy with Battery

There are several ways that extra energy from a storage element and/or sustainable sources may be utilized by the power network. Solar energy may be added to the overall network resources as is done today albeit with the caveat that its availability is not certain (that is the reason why we assign a finite probability to its energy supply). In the example (Table 3 below) we interfaced a solar panel with a battery. The parameters are the same as for the previous section with the addition of having a solar cap of 10 energy units.

|  | WITH SOLAR ENERGY AND NO OPTIMIZATION | WITH SOLAR ENERGY AND OPTIMIZATION |
|---|---|---|
| Energy distributed per round: | 99.53 | 99.978 |
| Energy requested per round: | 189.80 | 148.23 |
| Number of customers in the queue per round: | 92.29 | 96.73 |
| Number of customers that received energy per round: | 203.10 | 200.356 |
| Number of customers that requested energy per round: | 295.4 | 297.11 |
| Number of customers that were satisfied in the queue per round: | 4.83 | 47.73 |
|  |  |  |
| Energy distributed by the solar energy system per round: | 2.417 | 2.5571 |
| Energy available in the battery of the solar energy system per round: | 0.897 | 0.0020 |
| Solar energy produced per round: | 2.443 | 2.5571 |
| Number of customers that requested solar energy per round: | 9.312 | 49.9740 |
| Number of customers that received solar energy per round: | 2.476 | 6.49 |
| Total energy delivered (Solar+Grid) per round: | 101.9 | 102.55 |
|  |  |  |
| Number of rounds a customer in the queue waits to be satisfied: | 10.56 | 2.0856 |
| Number of rounds a customer spends in the queue | 9.229 | 9.67 |
|  |  |  |
| Wait time to receive energy from the solar system: | 0.2033 | 6.3794 |

*Table 3. A power network interfaced with a power storage): the comparison is made with and without optimization).* ***The main advantage for the optimized solution is the ability to accommodate more users in the queue and the decrease in the queueing time.*** *The time is measured in cycles (rounds, or time slots). 'Energy requested per round' includes new and queued energy requests – it becomes smaller for the overall optimized solution.*

## II.4. Optimization Scenarios – The Path of Energy Flow

Optimization of the power flow should not only include time but also space (path). In Figure 8 we present the statistics of connecting several alternative sources to several users using the IEEE 39 test bus system. The bus is made of sources (green nodes), users who receive energy (orange nodes), users who do not ask for energy (light yellow nodes), users who are in queue (red nodes), path-through users that do not tap into the energy flowing through them (blue nodes) and energy flow paths (light blue arrows). The program searches for the minimal path (the Dijkstra's method) to determine which source will be used to which user. There was no limit on the source or the path capacity however there was a limit of 5 users per source. Extra users were directed to another source nearby. The total energy delivered to the users cannot be larger than the global system capacity. For simplicity, an average energy loss of 6% per path was considered.

The statistics changes at each request cycle (round, or time slot) and we present here a snapshot of a randomly chosen cycle. For simplicity, we included the probability of connecting the nodes but not their associated loss. As we can see from the figure, the scenario is rather complex; some nodes forward energy, others consume it, and some nodes play several roles, such as generators and consumers.

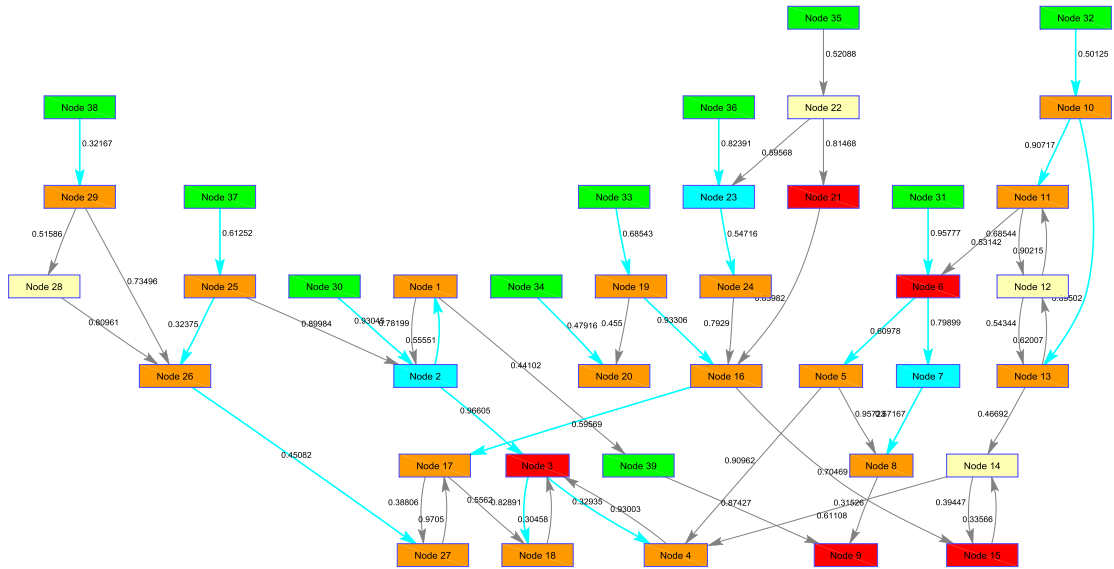

(a)

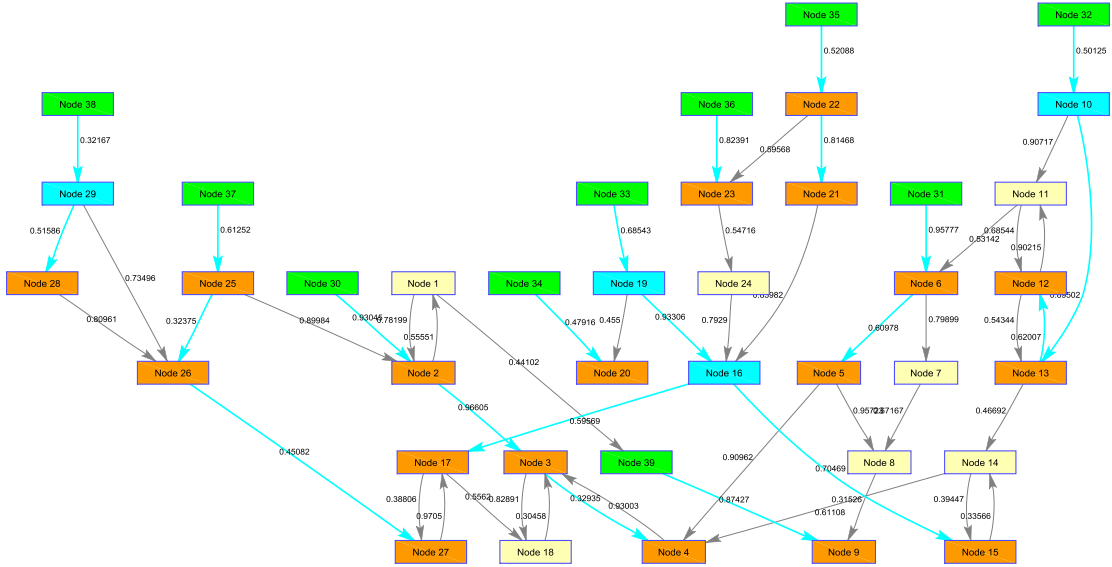

(b)

*Figure 8. (a) Simulating energy distribution in the IEEE 39 bus system. This is a snapshot at some particular round. Each number assigned to a dark gray arrow is the probability of a connected path. Red nodes: Users waiting in the queue; Green nodes: Sources; Light Yellow nodes: Users not requesting energy; Orange nodes: Users receiving energy; Blue arrows: Energy Path; Blue nodes: Users where energy is flowing through without tapping into it. (b) A situation where no node is in the queue*

## II.5. Realizing the DPN

The simplest system, which was a subject of a recent report is to have two overlaid networks [11]: one network provides the data and switching information and the other provides the energy (see Figure 1b). The energy in this case was delivered through existing transmission lines and in an analog format as provided today. The system was equipped with digital power switches (110 V, 15 A) and was dubbed ***Controlled Digital Grid (CDG)*** [9-11]. The difference between this system and a typical sensor network (also called the smart grid) is the incorporation of (active) power switches that open the path for the energy flow. With a request-grant protocol, a server receives the energy requests (power level and duration of service) upon a flip of a switch and communicates it to the electronic power switches along the path. Power limiters ensure that the power consumption does not exceed the requested limits. Alternative energy resources, as well as batteries are treated as users/sources depending on their function (Sections II.2. and II.3.)

One may also use the existing transmission lines for both data transmission and delivery of power. Data networks operate at much higher frequency rates than the power grid and the reactive elements of the power grid may be viewed as short for these frequencies. Several communication systems offer such service.

Digital energy may be realized by various ways (multiple frequencies; multiple phases; varying time slots; multiple current levels). Some of these maybe incompatible with the presently deployed grid (e.g., multiple frequencies may jeopardize the careful phase synchronization used today). Nevertheless, the key here is in the fusion of data and energy packets so that power switches are opened along the energy flow and no other user is able to interfere with that process. Such system is yet to be developed but may be applied at the smaller scale of micro-grids (e.g., the smart home).

## Summary

We have simulated the statistical behavior of Digital Power Networks with randomly requested energy packets. The advantage of a digitally coded energy lies in its delivery to specific addresses through a specific paths. The DPN can then optimize the distribution of energy during each request cycle. The DPN enables queueing of unsatisfied energy requests and handles them in a prioritized manner. Genetic algorithms were found useful in handling the queued requests at the expense of extra processing time.

## Acknowledgement

This work was funded in part by NSF Grant CNS-1641033.